# Repulsion-attraction switching of nematic colloids formed by liquid crystal dispersions of polygonal prisms

B. Senyuk,[a] Q. Liu,[a] P. D. Nystrom[a] and I. I. Smalyukh*[abcd]

[a]*Dept. of Physics and Soft Materials Research Center, Univ. of Colorado at Boulder, Boulder, Colorado 80309, USA.*

[b]*Materials Science and Engineering Program, Univ. of Colorado at Boulder, Boulder, Colorado 80309, USA.*

[c]*Dept. of Electrical, Computer, and Energy Engineering, Univ. of Colorado at Boulder, Boulder, Colorado 80309, USA.*

[d]*Renewable and Sustainable Energy Institute, National Renewable Energy Laboratory and Univ. of Colorado at Boulder, Boulder, Colorado 80309, USA.*

[*]*E-mail: ivan.smalyukh@colorado.edu.*

## Abstract

Self-assembly of colloidal particles due to elastic interactions in nematic liquid crystals promises tunable composite materials and can be guided by exploiting surface functionalization, geometric shape and topology, though these means of controlling self-assembly remain limited. Here, we realize low-symmetry achiral and chiral elastic colloids in the nematic liquid crystals using colloidal polygonal concave and convex prisms. We show that the controlled pinning of disclinations at the prisms edges alters the symmetry of director distortions around the prisms and their orientation with respect to the far-field director. The controlled localization of the disclinations at the prism's edges significantly influences anisotropy of the diffusion properties of prisms dispersed in liquid crystals and allows one to modify their self-assembly. We show that elastic interactions between polygonal prisms can be switched between repulsive and attractive just by controlled re-pinning the disclinations at different edges using laser tweezers. Our findings demonstrate that elastic interactions between colloidal particles dispersed in nematic liquid crystals are sensitive to the topologically equivalent but geometrically rich controlled configurations of the particle-induced defects.

**Introduction**

Unexpected behaviour of elastic colloidal dipoles in nematic liquid crystals (LCs), first reported by Poulin et al.,[1] attracted a great deal of interest in properties of dispersions of colloids in LCs. This discovery rapidly led to the development of a new branch of soft matter physics known as LC colloids.[2,3] The interest in LC colloids was further enhanced by their potential uses in engineering artificial functional materials with unusual properties not encountered in nature[4] and the ensuing technological applications.[5-8] LC colloids often manifest unusual elastic interactions, which cannot be realized in colloidal systems with isotropic fluid hosts.[1] Colloidal particles dispersed in LCs tend to deform an otherwise homogeneous director field[9] **n(r)** around them, depending on their shape and size,[10-17] topology[18,19] and surface anchoring boundary conditions.[2,20-23] The resulting elastic distortions of **n(r)** propagate to large distances away from the colloidal particle surfaces and can be analysed in terms of elastic multipoles, in analogy to their electrostatic counterparts[1,2,21] and as in nematostatics.[24,25] Colloidal interactions mediated by LC elasticity, as well as their self-assembly, were mostly studied for spherical particles, which typically create **n(r)**-deformations having a rotational symmetry with respect to a far-field director $\mathbf{n}_0$.[2,21-28] However, **n(r)**-deformations and topological defects around faceted colloidal particles with sharp edges were shown to be more diverse and complex.[29-35] Changing the propagation of defect lines at the edges of faceted colloidal particles can induce not only achiral director structures but also the ones with chiral symmetry of **n(r)** of the ensuing LC colloids.[16,25,30,31]

In this article, we realise low-symmetry achiral and chiral **n(r)**–dressed colloids in the nematic liquid crystals using polygonal concave and convex prisms with sharp edges. Although the symmetry of the solid particles themselves is rather high, the particle-induced defects tend to lower the symmetry of resulting LC colloids, which then depends on the configuration of defects accompanying them. We show that the laser-controlled pinning of the disclinations at the pre-selected prisms edges changes symmetry of the director distortions around the prisms, as well as their orientation with respect to the far-field director. We study how the anisotropy of Brownian motion of polygonal prisms dispersed in LCs depends on the configurational states of the disclinations pinned along the various prism's edges, which enables different forms of their reconfigurable self-assembly. We demonstrate that elastic interactions between polygonal prisms can be selectively reversed from repulsive to attractive just by re-pinning the disclinations at different edges using laser tweezers. Our findings demonstrate that elastic interactions between colloidal particles dispersed in nematic LCs are determined not only by the symmetry of the particles and their surface functionalization, but also by the detailed configurational states of disclination defects stabilized by the interaction of the particle-induced defects with the particle's geometric features, like edges between their faces.

**Materials and experimental techniques**

Polygonal colloidal particles were fabricated from silica ($SiO_2$) using direct writing laser photolithography.[11,16,18] First, a 1 μm thick layer of silica was deposited on a silicon wafer using plasma-enhanced chemical vapour deposition followed by spin-coating a layer of photoresist AZ5214 (from Clariant AG) on the top. The polygonal shapes were produced first in the photoresist layer by illumination at 405 nm with a direct laser-writing system DWL 66FS (Heidelberg Instruments) and then in the silica layer by inductively coupled plasma etching of uncovered $SiO_2$. The photoresist mask was then removed with acetone, exposing arrays of silica polygons on the top of the silicon wafer. To release polygonal prisms, the silicon substrate was etched using selective inductively coupled plasma. Following repeated washing and sonication, they were re-dispersed in deionized water. The resulting colloidal prisms with concave and convex polygonal bases were ~1 μm in thickness. By adjusting this procedure at the stage of direct laser writing, we fabricated colloidal particles in the forms of rhombic, concave pentagonal, hexagonal and convex pentagonal prisms. They all were treated with an aqueous solution (0.05 wt%) of N,N-dimethyl-N-octadecyl-3-aminopropyl-trimethoxysilyl chloride (DMOAP) to produce perpendicular boundary conditions and homeotropic alignment of the surrounding molecules of the room-temperature nematic LC 5CB (4-cyano-4'-pentylbiphenyl from Frinton Laboratories, Inc.) and nematic mixture E7 (EM Industries). Following the fabrication, the particles were re-dispersed in methanol. After mixing with a nematic LC and evaporation of methanol at ~70 °C overnight, the ensuing nematic dispersions were infiltrated into the cells assembled with two glass substrates separated by glass spacers defining the desired thickness $h$ of a cell gap, which was $h \approx 10$ and 16 μm in our experiments. Substrates were treated with DMOAP to obtain homeotropic LC cells with the far-field director orthogonal to the confining plates. To assemble planar LC cells with a uniform in-plane far-field director, the substrates were spin-coated with polyimide PI2555 (HD Microsystems) and then unidirectionally rubbed after polyimide cross-linking through baking at 270 °C for 1h. One of the two substrates utilized in the cell fabrication was only 150 μm thick to minimize spherical aberrations during imaging and optical manipulations experiments involving immersion oil objectives with a high numerical aperture (NA), while the other one was 1 mm in thickness.

Our studies utilize a multimodal experimental setup built around an inverted Olympus IX81 microscope, which we used for bright field and polarizing optical microscopy observations, for three-photon excitation fluorescence polarizing microscopy (3PEF-PM) and for optical manipulations with holographic laser tweezers. A tunable (680–1080 nm) Ti:sapphire oscillator (140 fs, 80 MHz, Chameleon Ultra-II, Coherent) was used as an excitation source for the 3PEF-PM imaging.[36] This three-dimensional imaging of **n(r)** was realized through the multi-photon-absorption-based excitation of cyanobiphenyl groups of 5CB or E7 molecules by using the femtosecond laser light at 870 nm; the resulting fluorescence signals were detected within a spectral range of 387–447 nm by a photomultiplier tube H5784-20 (Hamamatsu).[36] A mirror scanning unit FV300 (Olympus) was used to control an in-plane position of the focused excitation beam and its polarization was varied using a half-wave plate mounted immediately before a 100× (NA = 1.42) oil immersion objective. Optical manipulations of polygonal prism particles and defects around them were performed using a holographic optical trapping system operating at a wavelength of $\lambda$ = 1064 nm and described in details elsewhere.[36,37] The same objectives were used for both imaging and the holographic optical manipulations. Conventional polarizing and transmission-mode bright-field optical microscopy textures characterizing the LC director structures, colloidal polygonal prisms and their motion during elastic interactions were studied with a CCD camera (Flea, PointGrey), with the video microscopy data typically acquired at the 15 frames per second rate.

## Results

**Polygonal prisms with homeotropic anchoring in a nematic LC**

Orientation of complex-shaped colloidal particles in a LC host is determined by particles' shape and boundary conditions at their surface.[10,11,18] The point group symmetry of *m*-sided polygonal prisms is $D_{mh}$ and the rhombic prisms have $D_{2h}$ symmetry. Colloidal polygonal prisms with homeotropic boundary conditions are found most frequently oriented with their bases perpendicular to $\mathbf{n}_0$. In addition to Brownian motion, the elastic repulsion between the particles and cell substrates keeps these particles floating in bulk of LC, despite the action of gravity.[11,18,38] As an example, Figure 1a-f shows a rhombic prism in a homeotropic nematic cell; the particle has its base normal to $\mathbf{n}_0$ and parallel to the substrates. The variations of colours in polarizing microscopy textures obtained with a phase retardation plate (Fig. 1a), as well as the spatial distribution of fluorescence intensity analysed with respect to polarization of excitation light in 3PEF-PM images (Fig. 1d), are all consistent with the normal surface boundary conditions defining orientation of $\mathbf{n}(\mathbf{r})$ at the prism sides. This, incompatible with $\mathbf{n}_0$, orientation of the director at the prism sides causes director distortions (Fig. 1a,d) with a singular defect line. The particle-induced half-integer disclination forms a closed loop around the prism (Fig. 1e,f).[11,16,18] The local structure of the disclination is commonly of wedge type and can be characterized by the strength $k=\alpha/2\pi=-1/2$ ($\alpha$ is an angle by which the director rotates around the defect line's core as one circumnavigates it once[9]), similar to that often observed around spherical particles,[2,6] though this structure is altered as the line defect can jump between different geometric features of colloidal particles. Though the disclination's detailed structure is changing as it passes near different geometric features of the particle, its topology remains unaltered as such half-integer disclinations are the only topologically stable line defects in the bulk of nematic LCs. The closed loops of these half-integer defect lines are topologically equivalent to a point defect of elementary ±1 hedgehog charge compensating for the $\pm\chi/2=\pm1$ hedgehog charge of the colloidal particle[18] with the Euler characteristic $\chi=2$ and with perpendicular surface boundary conditions for $\mathbf{n}(\mathbf{r})$. Careful examination of the appearance of the prism in the textures taken with the bright field microscopy at different focus depths and for different particles (Fig. 1b,c) indicates that the defect line is localized away from the middle of the prism's sides and rather tends to be pinned near the horizontal edges of the prism, close to either top or bottom large-area faces (Fig. 1e,f). The placement of the defect line on the edges breaks mirror symmetry of $\mathbf{n}(\mathbf{r})$-distortions with respect to a horizontal plane normal to $\mathbf{n}_0$ and passing through the middle of the prism, thus inducing an elastic colloidal uniaxial[25] dipole moment $\mathbf{p}_\parallel$ normal to the prism's base and along $\mathbf{n}_0$. Although the distance dependencies of forces and potentials of inter-particle elastic interactions are strongly affected by confinement-induced screening, near-field effects, and the role of higher-order multipoles like quadrupoles, the dependence of forces between particles on orientation of the center-to-center separation vector is consistent with the conclusion that the lowest-order elastic multipole is of dipolar type. For example, when placed side-by-side, anti-parallel elastic colloidal dipoles attract while the parallel ones repel. Therefore, the point symmetry of the resulting elastic colloidal object formed by the particle and the surrounding $\mathbf{n}(\mathbf{r})$ and defect is $C_2$, which is lower than the symmetry of the prism itself.[24,25] Similar director distortions, with the half-integer defect line following the side faces near edges of the bottom or top large-area faces, tend to form around majority studied polygonal prisms (Fig. 2a-c), where the point group symmetry of the ensuing elastic colloidal objects formed by *m*-polygonal particles with large area faces orthogonal to the far-field director is $C_m$.

However, some of the prisms (roughly 20%) are observed to have their large-area faces slightly tilted with respect to $\mathbf{n}_0$ (Fig. 1g-I and Fig. 2d-f). Moreover, this tilting can be controlled by shifting fragments of the disclination loop up or down across the prism thickness using laser tweezers. One can further verify and characterize this prism's tilting by re-focusing the microscope on different sides of the prism. In bright field microscopy textures (Fig. 1g and Fig.2d-f), some of the sides of particles appear brighter, as in Fig. 1b, and some appear darker, as in Fig. 1c. These observations, along with 3PEF-PM imaging, allow us to conclude that the defect line changes its localization with respect to particle's geometry, jumping between the bottom and top large-area faces of the prism (Fig. 1h,i).

The disclination can change its vertical location by jumping either along vertical edges, which is the most common, (Fig. 1g,i) or anywhere along the prism sides, as shown in Fig. 2d,f. From now on, we will refer to defect lines pinned at one of the horizontal edges and defect lines changing their propagation between two horizontal edges as, respectively, "plain" and "kinky" disclinations or defect loops. While the polygonal prisms and elastic colloids with plain disclinations are achiral, LC colloidal objects with the kinky disclinations become chiral objects and are characterized with even lower point group symmetry, which is $C_1$ in the most general case (Fig. 1i). It induces a chiral biaxial dipole,[25] which in case of rhombic prisms is described by a longitudinal elastic moment $\mathbf{p}_\parallel$ along $\mathbf{n}_0$ and an elastic moment $\mathbf{p}_\perp$ in the plane normal to $\mathbf{n}_0$. The more complicated geometry of polygonal base with various configurations of defect lines may require isotropic, anisotropic, chiral and longitudinal dipoles[25] to fully describe resulting elastic interactions, though this is beyond the scope of the present study focused just on the demonstration that interactions can be switched. The nematic LC colloids with kinky disclinations can form spontaneously and also can be created from the elastic dipolar colloids with plain disclinations using optical tweezers. In the latter case, one can selectively change the vertical location of the disclination between the two horizontal edges at the opposite large-area faces by locally melting LC near prism sides to an isotropic phase with a high-power (100 mW and higher) trapping laser beam. This allows for a controllable tailoring of disclination's three-dimensional configuration and pinning at the edges of the colloidal prisms. In what follows, we discuss how the control of pinning of disclinations along edges of polygonal prisms can allow for pre-defining the symmetry of ensuing elastic colloids and their elasticity-mediated interactions and self-assembly.

**Brownian motion of colloidal prisms**

Brownian motion of colloidal particles in LCs is anisotropic with respect to $\mathbf{n}_0$ due to the intrinsic anisotropy of viscous properties of LCs.[26,28,39] The shape of colloidal particles and even the structure of the particle-induced defects can also contribute to this motion anisotropy.[20,40] Indeed, we observe this when probing how rhombic prisms undergo Brownian translational and rotational motion in the homeotropic LC cells. Using bright field optical video microscopy (Fig. 1b,g), we track the in-plane spatial position and orientation of the rhombic prisms and determine distributions of translational or rotational displacements $\Delta\mathbf{r}=\mathbf{r}(t+\tau)-\mathbf{r}(t)$ made by the particle from frame to frame over the elapsed time $\tau$,[39] where $\mathbf{r}$ to be substituted by $x$ or $y$ when measuring translational displacements in $x$ and $y$ directions and $\theta$ when measuring rotational displacements around $\mathbf{n}_0$. To determine the mean square displacement (MSD) $\langle\Delta\mathbf{r}^2\rangle$ of the rhombic prisms, we fit the experimentally obtained distributions with a Gaussian function in the form[39,41] $P(\Delta\mathbf{r}|\tau)=P_0(\tau)\exp[-\Delta\mathbf{r}^2/(2\langle\Delta\mathbf{r}^2\rangle)]$, where $P(\Delta\mathbf{r}|\tau)$ is the probability that a prism will displace by $\Delta\mathbf{r}$ over time $\tau$, $P_0(\tau)$ is a normalization constant. The width of the distribution is determined by $\langle\Delta\mathbf{r}^2\rangle=2D_r\tau$, where $D_r$ is a diffusion coefficient and index $r$ corresponds to the measured diffusion. The internal angles of a rhombic base of these particles are equal to 72° and 108°, which makes them elongated in one direction: longitudinal $a$ and transverse $b$ dimensions of the rhombic base are 14 and 10 μm, respectively (Fig. 1). Therefore, we take this shape anisotropy into account by converting the displacements $\Delta x_j$ and $\Delta y_j$ measured in the $x$-$y$ frame to displacements $\Delta a_j$ and $\Delta b_j$ relative to a body frame of the prism[40] $a$-$b$ rotating about $\mathbf{n}_0$ (Fig. 3a) using the following coordinate transformation:

$$\begin{bmatrix}\Delta a_j \\ \Delta b_j\end{bmatrix}=\begin{bmatrix}\cos\beta & \sin\beta \\ -\sin\beta & \cos\beta\end{bmatrix}\begin{bmatrix}\Delta x_j \\ \Delta y_j\end{bmatrix},$$

where $\beta$ is an angle of rotation of the body frame. These transformations reveal the effect of the particle shape anisotropy, showing that the mean square displacements along the longitudinal dimension of the prisms with a plain disclination are at maximum and along the transverse dimension are at minimum (Fig. 3a). The corresponding diffusion coefficients are found to be $D_a=13.5\times10^{-4}$ μm$^2$ s$^{-1}$ along the longitudinal and $D_b=7.3\times10^{-4}$ μm$^2$ s$^{-1}$ along the transverse direction. The anisotropy of Brownian motion measured for the prisms with a kinky disclination is different. While the maximum translational MSDs of the rhombic prisms with a plain disclination are along the longitudinal dimensions of the particle (Fig. 3a), interestingly, the maximum MSDs of the rhombic prisms with a kinky disclination are found to be along direction of the maximum out-of-plain tilt of the rhombic prism (Fig. 3c) caused by the jump of the disclination between horizontal edges. This reorientation of the dumbbell shape of MSDs can be forced by the proximity of edges of the tilted prism to the substrates. Both the orientation of the MSD angular diagram and its dumbbell shape (Fig. 3c) change significantly with the transformation of defects and tilting of the particles with respect to the far-field director. The corresponding maximum and minimum diffusion coefficients for a tilted prism were measured as $D_{max}=6.9\times10^{-3}$ μm$^2$ s$^{-1}$ and $D_{min}=1.4\times10^{-3}$ μm$^2$ s$^{-1}$, respectively. The details of the studied MSD anisotropy (Fig. 3a,c) change depending on the 3D configuration of a disclination loop pinned along particle's edges, as well as the ensuing tilting of the prism with respect to the far-field director. In both cases the anisotropy of diffusion for rhombic prisms with different disclinations has a dumbbell shape, which can be different for prisms with a different polygonal base. This behaviour can be understood as a result of the interplay of effects of particle shape and orientation with respect to the far-field director, as well as the induced director structure and the intrinsic LC's viscosity anisotropy.[11,16,26]

We also probed a rotational diffusion of the rhombic prisms dispersed in a homeotropic LC cell (Fig. 3b,d). The histograms in Fig. 3b,d show the angular displacements of a longitudinal dimension $a$ acquired for ~5 min at the time steps $\tau$=67 ms. Fitting these distributions allows to find rotational diffusion coefficients for the rhombic prisms with plain and kinky disclinations, which are $D_\theta$=9.9×10$^{-6}$ rad s$^{-1}$ and $D_\theta$=14.7×10$^{-5}$ rad s$^{-1}$, respectively. As one can see, diffusion properties of rhombic prisms can be significantly altered, both qualitatively and quantitatively, depending on the details of how disclinations pin at the edges of the polygonal prisms. As diffusion coefficients are related to the corresponding viscous drag coefficients for particles moving in a fluid LC host, the observed strong anisotropy needs to be accounted for in the characterization of elastic pair interactions and self-assembly of prisms, as we will discuss in detail below. Our model system, with controlled defect configurations and particle orientations, may allow for the future systematic studies of the anisotropic colloidal particle diffusion in LCs, as well as exploiting how it can be pre-engineered by controlling defects.

**Controlled reversal of elastic forces between colloidal prisms**

Elastic interactions of polygonal colloidal particles are known to be highly anisotropic.[10,11] To show that controlled pinning of disclination defects can further dramatically affect the elastic interactions between colloidal prisms, we use laser tweezers and bright field video microscopy (Figs. 4 and 5). The prisms are released nearby each other with the help of optical tweezers and their relative positions are then tracked with video microscopy (Fig. 4a). At certain distances, elastic interaction forces make particles "feel" each other, so that their motion becomes very different from the random Brownian jiggling because of being significantly affected by the direction and strength of the ensuing elastic inter-particle forces. The time dependent center-to-center separation $S(t)$ between the particles can be constructed from the tracking data (Fig. 4b), which then is used to calculate the velocity $dS/dt$ of these particles. The Reynolds numbers are small in our experiments, therefore the inertia effects can be neglected and the force of elastic interactions $F_{s\text{-}s}$ can be calculated from balancing it with the drag force $F_{s\text{-}s} \approx -c_f dS/dt$ exerted on the moving particles. For this study, the friction coefficient $c_f$ is calculated by using the Einstein relation and the diffusion coefficients determined as described above.[1,10,11,26,28,42] For evaluation of elastic interaction forces $F_{s\text{-}s}$, $c_f=k_BT/D_{av}$ was calculated using an average diffusion coefficient found as[10] $D_{av}=(D_a+D_b)/2$. In the homeotropic cells, polygonal prisms with a plain disclination attract when placed side to side (Fig. 4a,b) with their dipole moments $\mathbf{p}_\parallel$ are anti-parallel and repel when their $\mathbf{p}_\parallel$ moments are parallel (Fig. 5a,b). Sometime rhombic prisms' sides are slightly shifted with respect to each other due to the small (5-10°) rotation of the prisms right before the contact and this shift depends on the particle's shape, which may mean that the near-field interaction force is not exactly along the center-to-center separation vector. However, this behavior is highly dependent on the particle's shape and can be attributed to a combination of near-field effects and the role played by the higher-order elastic multipoles. For example, hexagonal prisms do not shift with respect to each other (Fig. 4c-e). The strength of these interactions is characterized in Figs. 4b and 5d. The elastic force characterizing the side-to-side elastic interactions is found to approach about 10 pN (the inset of Fig. 4b), which allows to estimate the corresponding maximum pair interactions potential (binding energy) of ~20,000$k_B T$, where $k_B$ is the Boltzmann constant and $T$ is temperature. The side-to-side attraction can result in numerous different forms of self-assembly of polygonal prisms (Fig. 4c-e), also dependent on the number $m$ of sides that the interacting prisms have.

We also use the rhombic prisms in homeotropic cells to study the effect of different configurations and pinning of disclinations on elastic inter-particle interactions. As already discussed above, the rhombic prisms with plain disclinations and anti-parallel $p_\parallel$ attract side-to-side (Fig. 4a) but repel when their $p_\parallel$ are parallel (Fig. 5a). We have selected one of the two repelling rhombic prisms with a plain disclination (Fig. 5a) and, using optical tweezers, locally melted LC at one side of the rhombic prism when the laser was ON and then quickly quenched LC back to the nematic phase when the laser was turned OFF. This, along with the refocusing the laser beam during quenching, allowed for changing the spatial localization of pinning of the disclination at this particular side from one horizontal edge to another, thus transforming director configuration from that with a plain defect loop to a kinky disclination loop (Figs. 1g-I and 5c) and creating the additional dipolar moment $p_\perp$. As a result, these two prisms start to attract each other when the "treated" side of one prism faces another untreated prism (Fig. 5c,d). Although elastic dipole moments both parallel and perpendicular to $n_0$ can be associated with the colloidal prism that has a kink in the disinclination loop (Fig. 5c), only $p_\parallel$ is present in the other colloidal prism undergoing pair interactions, which is due to the lack of symmetry breaking along directions orthogonal to $n_0$. The observed interactions, both repulsive and attractive, are natural when considering "elastic charges" associated with local tilting of director that accompanies colloidal prisms, though they cannot be easily analyzed in terms of interactions expected for point dipoles because lateral dimensions of particles are comparable to the interaction distances. The maximum forces of the attractive elastic interactions between these two prisms were measured to be ~3-4 pN (the inset of Fig. 5d), with the corresponding estimated pair interaction potential ~10,000$k_BT$. At the same time if these two prisms face each other with their untreated sides, they still repel. This demonstrates the profound effect the pinning of disclinations along the prism edges has on elastic interactions between particles. The controlled pinning of disclinations along edges can switch elastic interactions from repulsive (Fig. 5a,b) to attractive (Fig. 5c,d), which allows for the laser-guided optically reconfigurable self-assembly of particles. Since the sample thickness $h \approx 10$ μm was comparable to the particle's lateral dimensions and interaction distances, the strong surface boundary conditions for the director alignment at confining substrates result in screening of the long-range elastic interactions, so that the decay of elastic forces with distance is different from that expected on the basis of elastic multipole expansion.

**Formation and diffusion of colloidal assemblies**

Unlike LC colloids formed by truncated pyramids, which we studied previously,[11] all colloidal prisms in our present study have mirror symmetry planes passing through their midplanes. Therefore, the geometry of particles does not pre-select the localization of disclination loops at the top or bottom large-area faces, but rather makes these two possible scenarios equally probable. In the process of LC colloidal sample preparation and quenching of the LC from isotropic to nematic phase, we spontaneously obtain particles with either plain or kinky disclinations. However, laser tweezers allow for controlling localization of these defects with respect to the geometric features of particles, so that the elasticity-mediated self-assembly can be well controlled too (Fig. 6). For example, polygonal prisms in planar LC cells can be laser-treated so that they all have plain disclination loops defining orientations of elastic dipoles to exhibit either the side-to-side or base-to-base assemblies (Fig. 6a,b). For example, the hexagonal prisms with a plain disclination interact attractively side-to-side when their **p** are anti-parallel (Fig. 6a) or face-to-face when their **p** are parallel (Fig. 6b,c). The example shown for the latter case (Fig. 6b-d) demonstrates that such laser-assisted control of defects of the LC colloids enables assembly of many particles, so that the length of stacks of self-assembled thin prisms can be larger than their lateral dimensions. Since the particles can be added one-by-one, so that the geometry of the self-assembled stack is tuned, it is of interest to explore how diffusion of colloidal objects in LCs depends on their geometry.

Diffusion properties of our LC colloids are strongly anisotropic due to anisotropic viscous properties of the host LC fluid, anisotropic shape of the particles and the spatial configuration and pinning of the particle-induced defects (Fig. 3). Diffusion of a single hexagonal prism due to the Brownian motion is anisotropic with respect to the far-field director with diffusion coefficients along $D_{\parallel}=6\times10^{-4}$ µm$^2$ s$^{-1}$ and perpendicular $D_{\perp}=16.1\times10^{-4}$ µm$^2$ s$^{-1}$ to **n**$_0$. This strong anisotropy is caused by the anisotropic shape of the prisms aligned with respect to **n**$_0$. However, the ratio $D_{\parallel}/D_{\perp}$ changes when additional prisms attract base-to-base and form a chain assembly aligned along the far-field director (Fig. 6c-e). Figure 6f shows that coefficient $D_{\perp}$ significantly decreases when the number of prisms in the chain increases. On the other hand, diffusion of the assembly $D_{\parallel}$ along the director is almost independent of the number of prisms in the chain (Fig. 6f). This behaviour emerges from the effective resistance of the LC liquid to the prism moving along **n**$_0$, which is determined by the area of the base and stays unchanged with increasing the number of the prisms in the chain. In contrast, the resistance to the sidewise motion in directions orthogonal to **n**$_0$ depends on the length of the chain assembly and increases with increasing the number of the prisms N, so that the diffusion coefficient $D_{\perp}$ decreases inversely proportional to N and can be fitted with a function[43,44] $D_{\perp} \propto A/N$, where A is a fitting parameter (Fig. 6f). This analysis of how $D_{\perp}$ behaves with the growth of this colloidal assembly is only approximate as diffusion of such assemblies is further influenced by particle-induced defects and elastic distortions that are partly shared within the colloidal assembly. Interestingly, because of the LC's anisotropy of viscous properties, the ratio $D_{\parallel}/D_{\perp}$ increases above unity even for the colloidal dimer assemblies (Fig. 6f).

**Discussion**

We have demonstrated that topological defect lines can mediate reconfigurable elastic self-assembly of LC colloids. Previous studies demonstrated how similar effects can arise because of entangling the particles by disclination loops and networks,[45] as well as because of the localization of defect loops at bases of truncated pyramids, which were pre-engineered for defining various types of colloidal self-assembly, ranging from crystals to quasicrystals.[11] In the present case of prisms, the laser-induced jumps of defect loops between the edges at opposite large-area faces transforms the side faces from "sticky" attractive to repulsive and back, so that the colloidal assemblies of arbitrary complexity can be built by utilizing prisms with different numbers of faces. The need of using laser tweezers limits the potential for technological applications of such LC-colloidal composites but the potential use of azobenzene-containing capping ligands[46] to functionalize colloidal prisms may enable the control of disclination jumping for many particles simultaneously. On a more fundamental side, our system is ideal for exploring how colloids can be used in organizing matter in controlled ways well beyond the traditional condensed matter themes, such as crystals, quasicrystals, and glasses, but rather complex structures designed for a particular desired function, like in the biological world. For example, a combination of side-by-side and out-of-plane attractive interactions, controlled by kinky disclinations, could yield three-dimensional colloidal assemblies mimicking different objects or follow various themes from topology and geometry (e.g. forming colloidal knots or rings that are not fabricated to have these shapes[12] but rather form such topologically complex configurations as the result of pre-programmed colloidal self-organization). Furthermore, in addition to the reconfigurable assembly that deals only with closed loops of defects around each prism, our preliminary studies also show the possibility of entangling multiple particles by topologically more complex defect structures shared by multiple partricles,[47] which will be explored elsewhere.

## Conclusions

To conclude, we have realized low-symmetry achiral and chiral elastic colloids in the nematic LCs formed by the polygonal concave and convex facetted prisms with edges. We have shown that pinning of the disclinations at the prism edges can alter the symmetry of the director distortions and defects induced around the prisms, as well as their orientation with respect to the far-field director. Structural configurations of the disclinations pinned along the prism's edges significantly influence anisotropy of the diffusion properties of polygonal prisms dispersed in LCs and their self-assembly. Elastic interactions between polygonal prisms can be switched between repulsive and attractive through re-pinning the disclinations at different edges using laser tweezers. This can lead to the new forms of reconfigurable self-assembly of prisms and other facetted particles into crystalline or quasi-crystalline lattices for a host of photonics and electro-optics applications. Since configurations of defect lines correlate with orientations of the studied colloidal prisms, in addition to optical switching of elastic interactions with laser tweezers, our approach may potentially be extended to electric and magnetic switching, where torques can be applied either directly to the colloidal particles[48] or to the LC director,[49] depending on the type of used colloids and LCs.

## Acknowledgements

We acknowledge discussions with A. Martinez and T. Lee. We also acknowledge the support of the US National Science Foundation Grant DMR-1420736.

## References

1   P. Poulin, H. Stark, T. Lubensky, D. Weitz, *Science*, 1997, **275**, 1770-1773.

2   H. Stark, *Phys. Rep.*, 2001, **351**, 387–474.
3   C. Blanc, D. Coursault, E. Lacaze, *Liq. Cryst. Rev.*, 2013, **1**, 83-109.
4   S. C. Glotzer and M. J. Solomon, *Nat. Mater.*, 2007, **6**, 557–562.
5   Q. Liu, Y. Cui, D. Gardner, X. Li, S. He, I. I. Smalyukh, *Nano Lett.*, 2010, **10**, 1347-1353.
6   I. Muševič, M. Škarabot, M. Humar, *J. Phys.: Condens. Matter*, 2011, **23**, 284112.
7   Q. Liu, Y. Yuan, and I. I. Smalyukh, *Nano Lett.*, 2014, **14**, 4071–4077.
8   Y. Zhang, Q. Liu, H. Mundoor, Y. Yuan and I. I. Smalyukh, *ACS Nano*, 2015, **9**, 3097-3108.
9   P. G. de Gennes and J. Prost, *The Physics of Liquid Crystals*, Oxford University Press, Inc., New York, 2nd edn, 1993.
10  C. P. Lapointe, T. G. Mason and I. I. Smalyukh, *Science*, 2009, **326**, 1083–1086.
11  B. Senyuk, Q. Liu, E. Bililign, P. D. Nystrom and I. I. Smalyukh, *Phys. Rev. E: Stat., Nonlinear, Soft Matter Phys.*, 2015, **91**, 040501(R).
12  A. Martinez, M. Ravnik, B. Lucero, R. Visvanathan, S. Žumer and I. I. Smalyukh, *Nat. Mater.*, 2014, **13**, 258–263.
13  M. V. Rasna, K. P. Zuhail, U. V. Ramudu, R. Chandrasekar, J. Dontabhaktuni, S. Dhara, *Soft Matter*, 2015, **11**, 7674-7679.
14  M. V. Rasna, U. V. Ramudu, R. Chandrasekar, and S. Dhara, *Phys. Rev. E: Stat., Nonlinear, Soft Matter Phys*, 2017, **95**, 012710.
15  S. Hernàndez-Navarro, P.Tierno, J. Ignés-Mullol and F. Sagués, *Soft Matter*, 2011, **7**, 5109-5112.
16  B. Senyuk, M. B. Pandey, Q. Liu, M. Tasinkevych and I. I. Smalyukh, *Soft Matter*, 2015, **11**, 8758-8767.
17  S. M. Hashemi, U. Jagodič, M. R. Mozaffari, M. R. Ejtehadi, I. Muševič, M. Ravnik, *Nat. Commun.*, 2017, **8**, 14026.
18  B. Senyuk, Q. Liu, S. He, R. D. Kamien, R. B. Kusner, T. C. Lubensky and I. I. Smalyukh, *Nature*, 2013, **493**, 200-205.
19  Q. Liu, B. Senyuk, M. Tasinkevych and I. I. Smalyukh, *Proc. Natl. Acad. Sci. U. S. A.*, 2013, **110**, 9231–9236.
20  F. Brochard, P. G. de Gennes, J. Phys. (Paris), 1970, **31**, 691-708.
21  T. C. Lubensky, D. Pettey, N. Currier and H. Stark, *Phys. Rev. E: Stat., Nonlinear, Soft Matter Phys.*, 1998, **57**, 610–625.
22  B. Senyuk, O. Puls, O. M. Tovkach, S. B. Chernyshuk and I. I. Smalyukh, *Nat. Commun.*, 2016, **7**, 10659 (2016).
23  M. A. Gharbi, D. Seč, T. Lopez-Leon, M. Nobili, M. Ravnik, S. Žumer and C. Blanc, *Soft Matter*, 2013, **9**, 6911-6920.
24  V. M. Pergamenshchik and V. O. Uzunova, *Eur. Phys. J. E*, 2007, **23**, 161-174.
25  V. M. Pergamenshchik and V. A. Uzhunova, *Phys. Rev. E: Stat., Nonlinear, Soft Matter Phys.*, 2011, **83**, 021701.
26  H. Stark, D. Ventzki, *Phys. Rev. E: Stat., Nonlinear, Soft Matter Phys.*, 2001, **64**, 031711.
27  B. I. Lev, S. B. Chernyshuk, P. M. Tomchuk, H. Yokoyama, *Phys. Rev. E: Stat., Nonlinear, Soft Matter Phys.*, 2002, **65**, 021709.

28  M. Škarabot, M. Ravnik, S. Žumer, U. Tkalec, I. Poberaj, D. Babič, N. Osterman, and I. Muševič, *Phys. Rev. E: Stat., Nonlinear, Soft Matter Phys*., 2007, **76**, 051406.

29  F. Hung and S. Bale, *Mol. Simul*., 2009, **35**, 822-834.

30  P. M. Phillips, N. Mei, L. Reven, and A. D. Rey, Soft Matter, 2011, **7**, 8592-8604.

31  P. M. Phillips, N. Mei, E. R. Soulé, L. Reven, and A. D. Rey, *Langmuir*, 2011, **27**, 13335-13341.

32  S. M. Hashemi and M. R. Ejtehadi, *Phys. Rev. E: Stat., Nonlinear, Soft Matter Phys.,* 2015, **91**, 012503.

33  D. A. Beller, M. A. Gharbi, and I. B. Liu, *Soft Matter*, 2015, **11**, 1078-1086.

34  M. Cavallaro, Jr., M. A. Gharbi, D. A. Beller, S. Čopar, Z. Shi, T. Baumgart, S. Yang, R. D. Kamien, and K. J. Stebe, *Proc. Natl. Acad. Sci. U.S.A.*, 2013, **110**, 18804-18808.

35  B. Senyuk, Q. Liu, Y. Yuan, and I. I. Smalyukh, *Phys. Rev. E: Stat., Nonlinear, Soft Matter Phys.*, 2016, **93**, 062704.

36  T. Lee, R. P. Trivedi and I. I. Smalyukh, *Opt. Lett*., 2010, **35**, 3447–3449.

37  R. P. Trivedi, D. Engström and I. I. Smalyukh, *J. Opt*., 2011, **13**, 044001.

38  O. P. Pishnyak, S. Tang, J. R. Kelly, S.V. Shiyanovskii, and O. D. Lavrentovich, *Phys. Rev. Lett*., 2007, **99**, 127802.

39  J. C. Loudet, P. Hanusse, P. Poulin, *Science*, 2004, **306**, 1525.

40  A. Han, A. M. Alsayed, M. Nobili, J. Zhang, T. C. Lubensky and A. G. Yodh, *Science*, 2006, **314**, 626-630.

41  P. M. Chaikin, T. C. Lubensky, *Principles of Condensed Matter Physics* (Cambridge Univ. Press, Cambridge, 1995).

42  I. I. Smalyukh, A. V. Kachynski, A. N. Kuzmin, and P. N. Prasad, *Proc. Natl. Acad. Sci. U. S. A*., 2006, **103**, 18048-18053.

43  R. Clift, J. R. Grace, and M. E. Weber, *Bubbles, Drops, and Particles*, Academic Press, Inc., New York, 6th edn, 1978.

44  R. D. Blevins, *Applied Fluid Dynamics Handbook*, Van Nostrand Reinhold, 1984.

45  S. Čopar and S. Žumer, *Phys. Rev. Lett*., 2011, **106**, 177801.

46  A. Martinez, H. C. Mireles, I. I. Smalyukh, *Proc. Natl. Acad. Sci. U. S. A*., 2011, **108**, 20891-20896.

47  T. Araki and H. Tanaka, *Phys. Rev. Lett*., 2006, **97**, 127801.

48  B. Senyuk, M. C. M. Varney, J. A. Lopez, S. Wang, N. Wu, and I. I. Smalyukh, Soft Matter, 2014, **10**, 6014-6023.

49  C. P. Lapointe, S. Hopkins, T. G. Mason, and I. I. Smalyukh, *Phys. Rev. Lett.*, 2010, **105**, 178301.

**Figures and Captions**

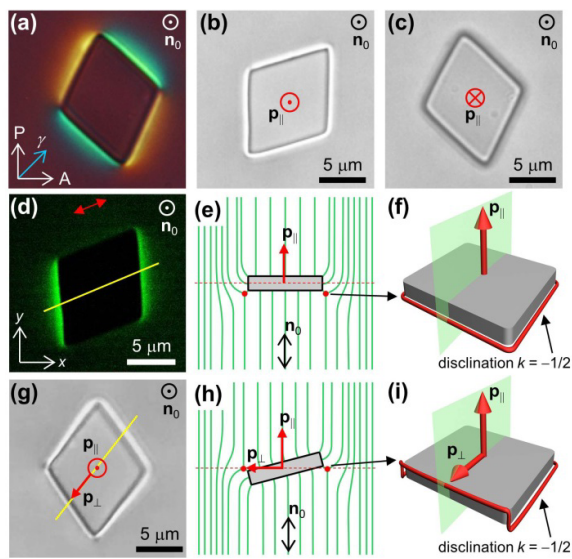

**Fig. 1** (a-d, g) Images of a rhombic prism with normal surface anchoring in a nematic homeotropic cell obtained with (a) polarizing, (d,c,g) bright-field and (d) 3PEF-PM optical microscopies. P, A and $\gamma$ mark, respectively, orientation directions of the crossed polarizer and analyzer and the slow axis of a 530 nm retardation plate inserted between the sample and analyzer at 45° to polarizers; the blueish and yellowish colours in a texture (a) indicate regions where $\mathbf{n}(\mathbf{r}) \| \gamma$ and $\mathbf{n}(\mathbf{r}) \perp \gamma$ respectively. Polarization of an excitation laser beam in the 3PEF-PM image shown in (d) is marked with a double red arrow. (e,f,h,i) Schematic diagrams showing the director configurations of $\mathbf{n}(\mathbf{r})$ (green lines) and the half-integer disclination (a red line forming the closed loop) induced around the polygonal prism in the planes marked in (d,g) and (f,i) by a yellow line and green plane, respectively. Black and white circle with a central dot in (a-d, g) and a double arrow in (e,h) show the far-field director $\mathbf{n}_0$. A red circle with a central dot or a cross in (b, c) and red arrows in (e-i) show the direction of the elastic moments $\mathbf{p}_\|$ and $\mathbf{p}_\perp$ of the colloidal dipoles, which originate from the asymmetric localization of disclination loops with respect to the midplanes of the prisms.

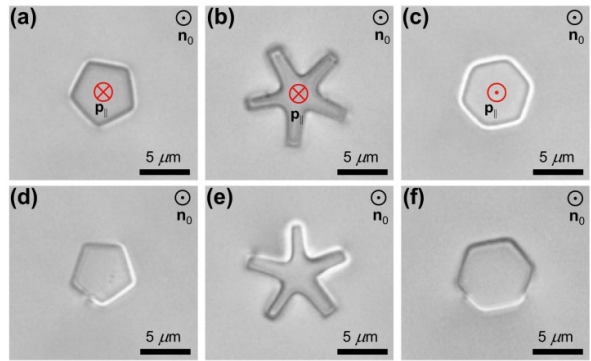

**Fig. 2** Bright-field optical micrographs of polygonal prisms (a-c) with disclination loops confined to single planes coinciding with the bottom or top large-area faces of the particles and (d-f) the same prisms after "jumping" of the disclination between the edges at the top and bottom faces was prompted using laser tweezers.

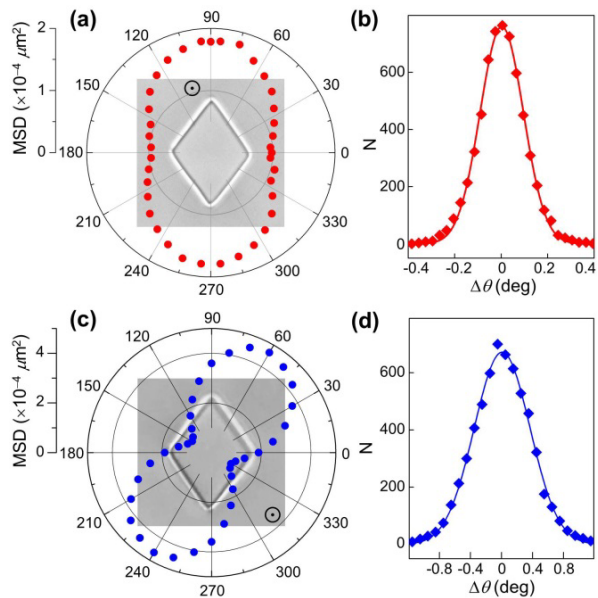

**Fig. 3** (a,c) Angular dependencies of a mean square translational displacement (MSD) and (b,d) angular displacements of a rhombic prism with (a,b) a plain disclination and (c,d) a kinky disclination in a homeotropic LC cell ($h \approx 10$ μm).

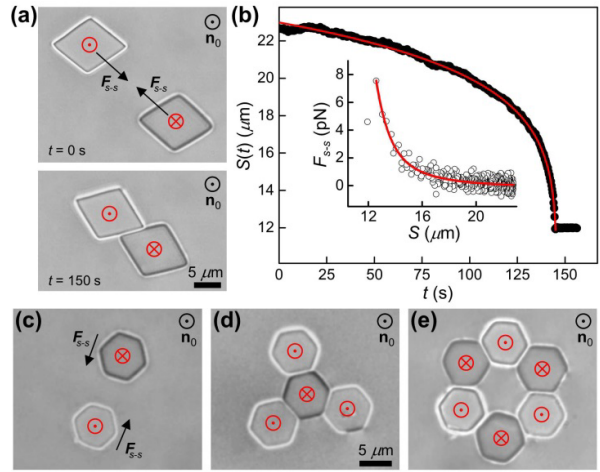

**Fig. 4** Side-to-side attractive elastic pair interactions between (a) rhombic and (c) hexagonal prisms in a homeotropic cell ($h \approx 10$ µm). (b) Center-to-center separation vs. time for attractive elastic interactions of (a) rhombic particles. The inset shows a dependence of the force on the separation between particles. Red lines are eye guides. (d,e) Examples of assemblies of hexagonal prisms enabled by dipole-like elastic interactions between particles with defect loops localized at different large-area top or bottom faces.

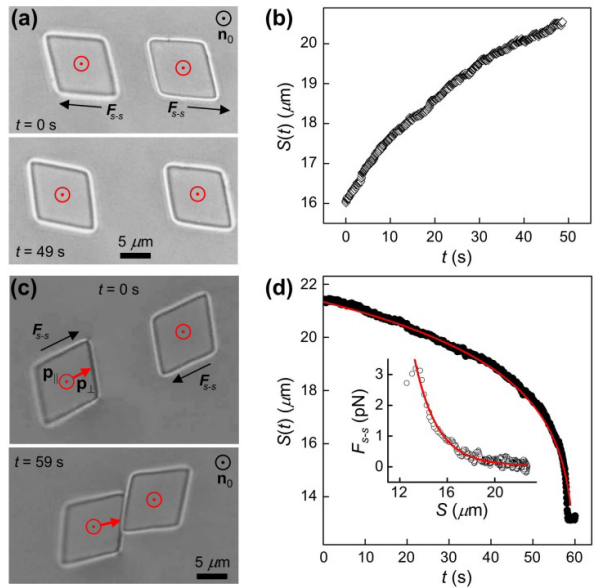

**Fig. 5** (a) Repulsive elastic pair interactions of rhombic prisms with a plain disclination and parallel elastic moments in a homeotropic cell ($h \approx 10$ µm) and (b) their center-to-center separation vs. time. (c) Attractive elastic pair interactions between rhombic prisms with a kinky (a prism on the left) and plain (a prism on the right) disclination and (d) their center-to-center separation decreasing with time. The inset in (d) shows a dependence of the force of interactions on the separation between particles.

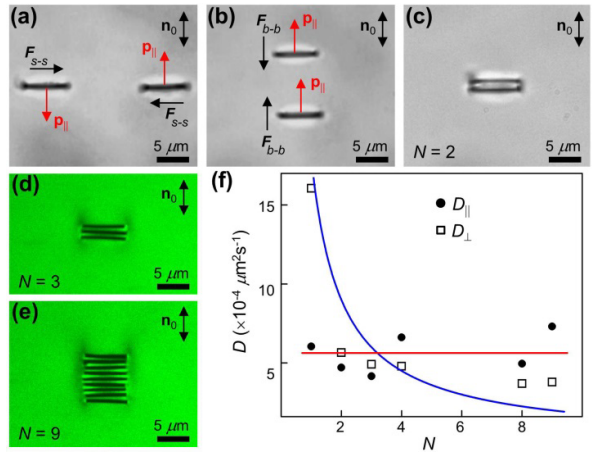

**Fig. 6** (a-c) Bright-field optical and (d,e) 3PEF-PM microscopy images showing (a) side-to-side and (b) base-to-base attraction of hexagonal prisms in a planar LC cell ($h \approx 16$ µm). The orientations of elastic dipole moments $\mathbf{p}_\parallel$ are shown using red arrows and the side-to-side and base-to-base force directions are marked with black arrows and "$F_{s\text{-}s}$" and "$F_{b\text{-}b}$", respectively. (c-e) Linear assemblies of respectively 2, 3 and 9 hexagons that result from the base-to-base dipolar elastic attraction between particles with parallel $\mathbf{p}_\parallel$, such as the ones shown in (b). (f) Diffusion coefficients of a linear assembly of hexagonal prism along ($D_\parallel$) and perpendicular ($D_\perp$) to $\mathbf{n}_0$ depending on the number $N$ of particles within the assembly. Red and blue lines show fitting of experimental data with the corresponding functions.